\documentclass{article}
\usepackage{spconf,amsmath,graphicx,hyperref}
\usepackage{graphicx}
\usepackage{tabularx}
\usepackage{multirow}
\usepackage{float}
\usepackage{booktabs}
\usepackage{amsfonts} 
\usepackage[table]{xcolor}
\ninept

\title{Break-the-Beat! Controllable MIDI-to-Drum Audio Synthesis}

\name{
\begin{tabular}{@{}c@{}}
Shuyang Cui$^{1}$,
\quad Zhi Zhong$^{1}$,
\quad Qiyu Wu$^{1}$,
\quad Zachary Novack$^{1*},$\thanks{$^*$Work done during the internship at Sony Group Corporation.}\\
Woosung Choi$^{2}$,
\quad Keisuke Toyama$^{1}$,
\quad Kin Wai Cheuk$^{2}$,
\quad Junghyun Koo$^{2}$,\\
Yukara Ikemiya$^{2}$,
\; Christian Simon$^{1}$,
\; Chihiro Nagashima$^{1}$,
\; Shusuke Takahashi$^{1}$
\end{tabular}}
\address{$^{1}$Sony Group Corporation \;
$^{2}$Sony AI\
}
\begin{document}
%
\maketitle

\begin{abstract}
Current methods for creating drum loop audio in digital music production, such as using one-shot samples or resampling, often demand non-trivial efforts of creators. While recent generative models achieve high fidelity and adhere to text, they lack the specific control needed for such a task. Existing symbolic-to-audio research often focuses on single, tonal instruments, leaving the challenge of polyphonic, percussive drum synthesis unaddressed. We address this gap by introducing ``Break-the-Beat!,'' a model capable of rendering a drum MIDI with the timbre of a reference audio. 
It is built by fine-tuning a pre-trained text-to-audio model with our proposed content encoder and a effective hybrid conditioning mechanism. To enable this, we construct a new dataset of paired target-reference drum audio from existing drum audio datasets. Experiments demonstrate that our model generates high-quality drum audio that follows high-resolution drum MIDI, achieving strong performance across metrics of audio quality, rhythmic alignment, and beat continuity. This offer producers a new, controllable tool for creative production. Demo page: \url{https://ik4sumii.github.io/break-the-beat/}
\end{abstract}
\vspace{-2mm}
\section{Introduction}
\label{sec:intro}
\vspace{-2mm}

In digital music production, drums play a foundational role in shaping the rhythm, energy, and overall character of a composition. 
Conventional workflows for creating expressive drum mixes typically requires non-trivial efforts using Musical Instrument Digital Interface (MIDI). 
However, synthesizing high-quality drum mixes is challenging for creators as it requires both advanced mixing skills and the effort of creating one-shot samples.
Non-experts often produce drum mixes with discontinuous sounds or unintended tone.
In this context, a model capable of synthesizing drum audio that faithfully \textit{follows a given drum MIDI} (where each pitch corresponds to a specific drum group) while also reproducing the \textit{timbre of a reference audio} would be transformative for music production workflows. 

Recently, many music generative models have been proposed to assist creators in their workflow. 
For example, text-to-music models  have shown impressive fidelity \cite{evans2024stable, Novack2025Fast}.
To control generative models with a given melody, several attempts have been made \cite{wu2023music, Novack2024Ditto, Novack2024DITTO2DD}. 
In these works, Chroma, extracted from raw waveform, was adopted as a control signal.
Hou \textit{et al.} \cite{hou2024editing} used a pretrained Stable-Audio-Open (SAO) \cite{evans2024stable} to render CQT representation into realistic music signals, controlling timbre with text prompt. MuseControlLite \cite{tsai2025musecontrollite} used a similar idea but reduced the required parameters to achieve the control. 
The closest research filed to our interest would be MIDI-to-Piano, a task that synthesizes natural piano audio given MIDI conditions. 
Recent progress has been made in \cite{tang2025midivalle}, where the method is trained on large-scale piano datasets. 
Apart from these recent generative models, conventionally, transcription has been the focus of drum-related \cite{Vogl2017DrumTV,Callender2020ImprovingPQ, Torres2025TheID} and MIDI-related \cite{hawthorne2021sequence-piano-, gardner2021mt3, toyama2023hft} research. 
In summary, there is no prior work on MIDI-to-Drum generation, where MIDI serves as a control signal to synthesize a drum track.
Most existing works do not employ MIDI as control signal, with \cite{tang2025midivalle} being the exception.
However, while \cite{tang2025midivalle} targets piano generation, our goal is drum generation, where the timbral and rhythmic characteristics are fundamentally different.
Our contributions are as follows:
(1) We propose ``Break-the-Beat!,'' a Diffusion model built upon the pre-trained SAO \cite{evans2024stable}, to the best of our knowledge, a pioneer for MIDI-to-Drum synthesis;
(2) We provide design insights. We found a MIDI grid with higher resolution is beneficial to rhythm alignment. A proper combination of conditioning mechanisms contributes to both audio synthesis quality and rhythm alignment. Using pre-trained SAO is a key to achieve high-quality synthesis under limited training epochs.
(3) We provide an evaluation framework including qualitative and quantitative evaluations to facilitate the advancing of this research topic. 

\begin{figure*}[t]
    \vspace{-15mm}
    \centering    
    \includegraphics[width=0.9\textwidth]{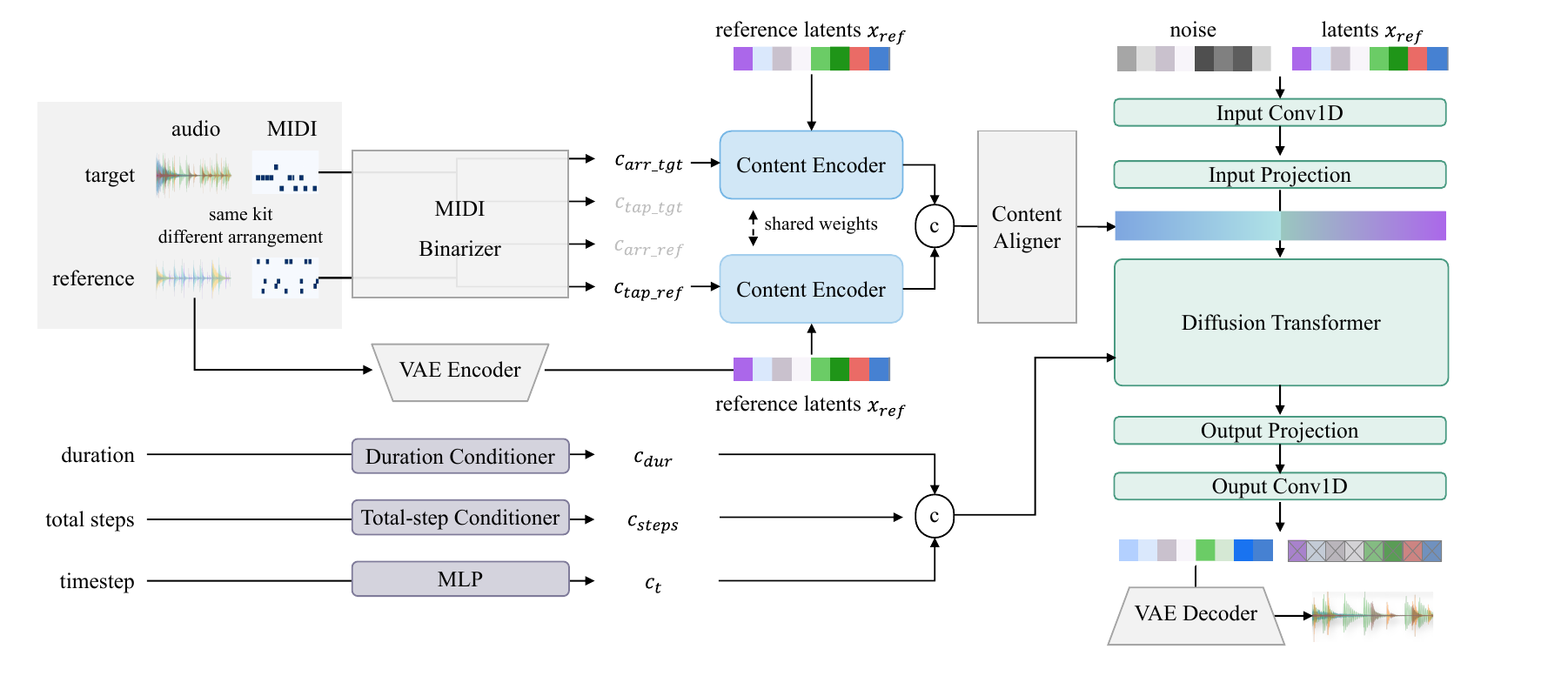}
    \vspace{-7mm}
    \caption{Overview of our proposed method. Stable Audio Open framework \cite{evans2024stable}, originally designed for text-to-audio generation, is adapted into a model conditioned on drum MIDI and reference audio.}    
    \label{fig:overview}
    \vspace{-5mm}
\end{figure*}

\vspace{-2mm}
\section{Related Work}
\vspace{-2mm}

Generative audio models have achieved impressive fidelity in recent years \cite{evans2024stable, Novack2025Fast, pmlr-v202-liu23f, Liu_2024, saito2024soundctm}. 
However, providing precise and expressive control to such generative models remains an open challenge, particularly in domains such as music production and sound design. These fields demand not only high fidelity but also the model's ability to adhere to specific creative intentions, such as defining exact rhythms, melodies, or timbres across single or multiple instruments. 

Early music generative models \cite{evans2024stable, pmlr-v202-liu23f, Liu_2024} allowed limited control, introduced latent diffusion for text-conditioned generation, but did not support precise event-level or temporal control. 
To improve the controllability and expressivity of generative model, recent works have incorporated fine temporal control on top of text conditions \cite{wu2023music, Novack2024Ditto, Novack2024DITTO2DD, tsai2025musecontrollite, levy2023controllable}. 
Beyond text-to-audio generation with temporal conditions, another line of research focuses on modeling the transformation of symbolic content into realistic audio waveforms. 
This approach aims to bridge the gap between abstract symbolic representations, such as MIDI sequences \cite{tang2025midivalle, zhang2025renderbox, tang2025pianovits, Kim2025TokenSynthAT} or rhythmic patterns, and high-fidelity audio wave forms. 
A well-studied analogous problem exists in text-to-speech synthesis, where prior works have successfully modeled the conversion of linguistic content into natural speech while also preserving the voice characteristics of a source speaker by means such as auto-regression \cite{defossez2024moshi, ye2025llasa}, MaskGIT \cite{borsos2023soundstorm, wang2024maskgct}, or a hybrid of them \cite{wang2023valle, neekhara2024t5tts}. More recently, Diffusion models \cite{eskimez2024e2tt2, chen2024f5tts} have been introduced to TTS tasks.
Similarly, one could imagine a ``speech of drum'' system that renders a given rhythmic sequence with the sound of a different drum kit. This highlights the need for a generative model capable of separately controlling ``what is played'' (the drum pattern) and ``how it sound'' (the drum timbre) within a unified framework. 
However, previous symbolic-to-audio generation research mainly focuses on single-instrument tonal audio synthesis such as Piano, leaving the challenge of percussive, atonal midi-to-drum audio synthesis largely unaddressed.
\vspace{-3mm}
\section{Method}
\label{sec:method}
\vspace{-2mm}
Fig. 1 shows the overview of our proposed method. We utilize the Stable Audio Open (SAO) framework \cite{evans2024stable}, which incorporates the Diffusion Transformer (DiT) for text-to-audio generation. In our work, we adapt the DiT model conditioned on drum MIDI and reference audio.
We first describe the input representations (\S \ref{sec:Conditional Input Representations}) and our content encoder (\S \ref{sec:Dual-Input Content Encoder}) for conditioning signals. The conditioning mechanism for DiT is summarized in \S \ref{sec:Hybrid Conditioning Mechanism}, and the overall training procedure is outlined in \S \ref{sec:Training Strategy}. Lastly, we detail the preparation of the training dataset in \S \ref{sec:Training Data Formulation}. 

\vspace{-2mm}
\subsection{Conditional Input Representations}
\label{sec:Conditional Input Representations}
\vspace{-1mm}
We need to inject MIDI into SAO to control the rhythm, and we need a reference audio to extract timbre. We explain the representation for MIDI and reference audio in this section.

\noindent\textbf{Representation of Drum MIDI.} We follow \cite{gillick2019learning} in categorizing drums into nine instrument groups. Each MIDI note specifies a drum hit by note start time and pitch. Instead of absolute time grids, which entangle tempo and pattern, we use a relative encoding via a \textit{MIDI Binarizer}, quantizing the sequence to a fixed rhythm grid based on tempo, with a timestep resolution of n-th \footnote{e.g., 1/16 note: \url{https://en.wikipedia.org/wiki/Sixteenth_note}}. 
This results a sequence with $T$ steps, which is the maximum number of n-th notes. We examine the impact of choosing different timestep resolution in Table~\ref{table:granularity}. Each timestep is a 10-dimensional binary vector. The first nine dimensions indicate the instrument hit, while the 10th is a global onset indicator. 
We omit velocity and time-offset value, encouraging the model to learn the expressive details from data. 
This results in two types of representations: \textbf{Arrangement} $\mathbf{c}_\text{arr} \in \mathbb{R}^{T \times 10}$ (use first nine dimensions for instrument-specific arrangement), \textbf{Tap} representation $\mathbf{c}_\text{tap} \in \mathbb{R}^{T \times 10}$ (use the 10th dimension for onset), both decoupling rhythmic structure from absolute timing. We encourage readers to visit our demo page for samples generated from same MIDI but with different tempo.

\noindent\textbf{Representation of Reference Audio.} We leverage the pre-trained VAE encoder from \cite{evans2024stable}, which trained on a wide variety of audio, to encode the reference audio into a sequence of latent vectors, obtaining $x_\text{ref} \in \mathbb{R}^{N_\text{ref}\times d_\text{aud}}$, where $N_\text{ref}$ is the number of the latent frames and $d_\text{aud}=64$ is the latent dimension. Assuming these latent vectors effectively encode the timbral and temporal characteristics of the reference audio, employing them as control signals allows the model to synthesize drum audio that follows the MIDI rhythm and adopts the reference timbre, enabling flexible timbre control beyond the training data.

\vspace{-2mm}
\subsection{Dual-Input Content Encoder}
\label{sec:Dual-Input Content Encoder}
\vspace{-1mm}
Our content encoder, denoted as $f_{\text{enc}}^\theta$, is designed to process the representations described in \ref{sec:Conditional Input Representations}. 
It consists of a multi-layer self-attention transformer encoder and a multi-layer cross-attention transformer encoder. The self-attention layers first capture the internal temporal structure within the input MIDI representations. Subsequently, the cross-attention layers take the output of the self-attention layers as the Query, and $x_\text{ref}$ as the Key and Value, thereby infusing the temporal features of MIDI with spectral and temporal structure of reference audio.

To provide the model with the understanding of both the target arrangement and the timbre within the reference audio, we employ a dual-input strategy, \textit{i.e.}, $f_{\text{enc}}^\theta$ processes two inputs in parallel with shared weights: (1) the target MIDI representation $\mathbf{c}_{\text{tgt}}$ and (2) the corresponding MIDI representation of the reference audio $\mathbf{c}_{\text{ref}}$.
The resulting arrangement feature sequences are then concatenated along the sequence dimension. This yields the final conditioning content features $\mathbf{c}_{\text{cont}} = [f_{\text{enc}}^\theta(\mathbf{c}_{\text{tgt}}, x_\text{ref}),f_{\text{enc}}^\theta(\mathbf{c}_{\text{ref}}, x_\text{ref})] \in \mathbb{R}^{2T \times d}$.
  
\vspace{-2mm}
\subsection{Hybrid Conditioning Mechanism}
\label{sec:Hybrid Conditioning Mechanism}
\vspace{-1mm}

We employ a hybrid mechanism to condition the DiT, utilizing three distinct methods:

\noindent\textbf{Concatenation.} The reference audio latents $x_{\text{ref}}$ are concatenated to the noisy latent sequence $z_t$, providing reference context for timbre.

\noindent\textbf{Input Addition.} The content features $\mathbf{c}_{\text{cont}}$ are first temporally aligned by a \textit{Content Aligner}. The aligned features are then added element-wise to the DiT's input after the initial pre-process 1D convolution layer inspired by \cite{garca2024sketch2sound}. To bridge the different temporal resolutions between the rhythm-grid-based content features $\mathbf{c}_{\text{cont}}$ and the time-based audio latents, we propose a simple, training-free \emph{Content Aligner}. We find the feature that is nearest in time from $\mathbf{c}_{\text{cont}}$ for each step of audio latents according to the tempo, and then copy each feature to the designated step. 


\noindent\textbf{Prepending.} Following~\cite{evans2024stable}, we encode the diffusion timestep $t$ ($\mathbf{c}_{t}$) with an MLP and the target duration ($\mathbf{c}_{\text{dur}}$) with a \textit{duration conditioner}. We extend this by also encoding the number of timestep ($\mathbf{c}_{\text{steps}}$) of target arrangement using another \textit{total-step conditioner} similar to \textit{duration conditioner}. All three embeddings of these global conditions are prepended to the DiT's input sequence.

The effectiveness of using this \textit{Hybrid Conditioning Mechanism} is empirically validated in Table \ref{table:ablation}.
\vspace{-2mm}
\subsection{Training Strategy}
\label{sec:Training Strategy}
\vspace{-1mm}
During training, the DiT is jointly optimized with content encoder and encoders that process global conditions. Following \cite{evans2024stable}, we use a sinuous noise scheduler to add noise to the clean latent variable of target audio $x_{tgt}$, producing noisy latent $z_t=\alpha_t \cdot z_0 +\sigma_t z_1 $, where $z_1 \sim \mathcal{N}(0, I)$ is sampled noise and $z_0=x_{tgt}$. The training objective is based on the v-objective \cite{salimans2022progressivedistillationfastsampling}:
\begin{equation}
\mathcal{L}_{\textbf{v-diffusion}}(\theta;z_t, t\mid Y) = {\left \| v_\theta(z_t,Y) - (\alpha_tz_1-\sigma_tz_0)\right \| }^2
\end{equation}
, where $Y=\left\{\mathbf{c}_{tgt}, \mathbf{c}_{ref}, x_{ref}, \mathbf{c}_{dur}, \mathbf{c}_{step}, \mathbf{c}_{t} \right\}$ is the set of conditions defined above, and $v_\theta$ is the model. We also apply \emph{classifier-free guidance} by blanking out $\mathbf{x}_{ref}$ with 10\% probability following \cite{evans2024stable}, forcing the model to learn an unconditional prediction.

We employ a curriculum training strategy that gradually increases task difficulty. For the target condition, we linearly shift the input probability from 100\% \textbf{Arrangement} at the start to an even 50/50 split with \textbf{Tap} as training progresses. Concurrently, for the reference condition, the input starts from a 50/50 split between its \textbf{Arrangement} and \textbf{Tap} representations to using the \textbf{Tap} only. This progression allows the model to first master the task of arrangement-to-audio synthesis before tackling the more ambiguous task of tap-to-audio synthesis, while also adapting it for the practical scenario where the reference audio lacks a ground-truth \textbf{Arrangement} and only has \textbf{Tap}.
\vspace{-2mm}

\subsection{Training Data Formulation}
\label{sec:Training Data Formulation}
\vspace{-1mm}
Our goal is to train a model that can synthesize a drum audio following a target arrangement or tap while utilizing the timbre of a given reference audio. To facilitate this specific task, we construct a paired dataset from existing drum recording datasets paired with MIDI, as no readily available dataset serves this purpose. We formulate the task by constructing reference-target pairs where, for each target drum audio, we pair it with a reference audio that is rendered using the same drum kit but a different MIDI sequence.  Furthermore, to allow the model to distinguish individual drum instruments and their combinations, our training data combines both full mixtures and isolated stems. 

\begin{figure*}[t]
    \centering
    \includegraphics[width=\textwidth]{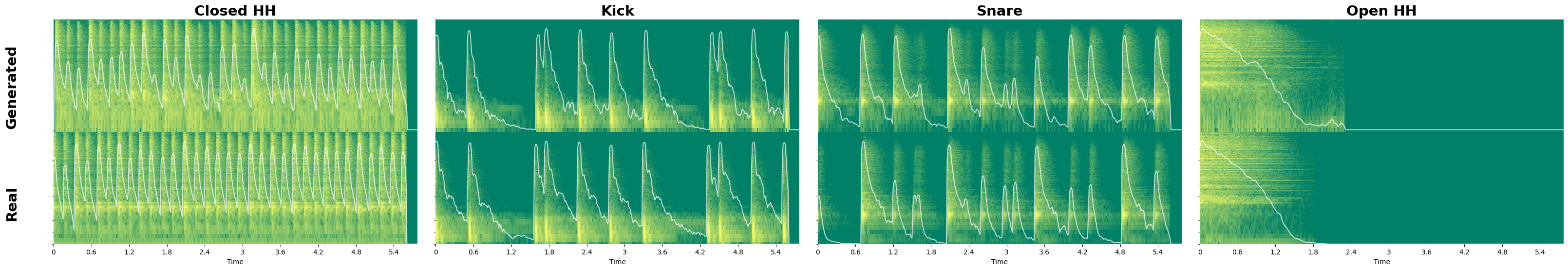}
    \vspace{-8mm}
    \caption{From left to right demonstrates results of synthesized instrument-wise drum audio from multi-drum arrangement in StemGMD, comparing with ground-truth.}
    \vspace{-3mm}
    \label{fig:stem}
\end{figure*}
\vspace{-2mm}
\section{Experiments}
\label{sec:experiments}
\vspace{-2mm}
\subsection{Data}
\vspace{-1mm}
We train and evaluate our approach on two variations of the Groove MIDI Dataset (GMD)\cite{gillick2019learning}, which consists of 1059 unique human-performed MIDI drum sequences aligned with corresponding audio recordings, where the vast majority ($\sim~99\%$) use a 4/4 time signature and a significant portion ($\sim~66\%$) are shorter than 10 seconds. The two derivatives are the \textit{Expanded Groove MIDI Dataset (E-GMD)} \cite{Callender2020ImprovingPQ}, contains audio recordings rendered with 43 diverse drum kit presets, and \textit{StemGMD} \cite{Mezza_2024}, which uses another 10 distinct drum kit presets but records isolated audio stems for each drum group. 

For all experiments, we extract a 2-bar segment (equivalent to 8 quarter notes in 4/4 time signature) from each audio-MIDI pair to train the model efficiently. This segment length is a typical length for drum loops as it effectively captures the core rhythmic variations in a drum beat while being suitable for evaluation \cite{gillick2019learning}. This pre-processing yields a total of 76.68 hours of audio samples. To analyze the effect of temporal granularity, we experiment with three timestep resolutions: 16th, 32nd, and 64th notes, for MIDI quantization. We follow the official train/validation/test splits for each dataset and further ensure that no MIDI files overlapped between the sets to prevent data leakage.

To construct reference-target pairs, we enforce a strict separation at the drum kit level: a subset of kits is reserved exclusively for the validation and test sets. This ensures that the model's generalization capability to unseen timbres is properly evaluated. This procedure results in a final dataset of 62,595 training pairs, 1,202 validation pairs, and 791 test pairs.

\vspace{-2mm}
\subsection{Model}
\vspace{-1mm}
We initialize our model with the pre-trained DiT and VAE checkpoints from SAO \cite{evans2024stable}. Our training fine-tunes the 24-layer DiT, keeping the VAE parameters frozen. We also experiment with training the DiT from scratch. The content encoder is a 4-layer model, following the design of DiT blocks. The stereo audio waveform is sampled at 44.1kHz and encoded by the pre-trained VAE encoder with a compression rate of 2048. We use the AdamW optimizer with learning rate of 1e-4 and InverseLR scheduler. We train the model on 8 H100 GPUs for 50 epochs with a batch size of 4 per GPU. For inference, we employ the DPM-Solver++ sampler \cite{lu2022dpmsolver} with 10 sampling steps.
\vspace{-7mm}
\subsection{Metrics}
\vspace{-1mm}
We evaluate the quality of synthesized drum audio by our models using the following three aspects. 

\noindent\textbf{Audio Quality}. To evaluate the overall perceptual quality of synthesized audio, we use Fréchet Audio Distance \cite{Kilgour2019FrchetAD} based on VGGish-16kHz \cite{Hershey2016CNNAF} and CLAP-LAION-48kHz  \cite{Wu2022LargeScaleCL} embeddings (checkpoint of music\_audioset\_epoch\_15\_esc\_90.14), as an approximate measure of drum timbre similarity due to the lack of a precise drum timbre metric.

\noindent\textbf{Alignment}. To quantify how accurately the hit of synthesized audio adheres to the ground-truth audio, we measure the onset-based F1-score. First, we detect onset timestamps in both audios using \texttt{librosa} \cite{McFee2015librosaAA}. An onset in the synthesized audio is considered a correct hit if it falls within a 100 ms tolerance window of an onset in the ground-truth audio. Additionally, we compute the RMS error in dBFS. This metric quantifies the similarity of the overall dynamic. 

\noindent\textbf{Beat Continuity}. To evaluate the temporal stability and rhythmic coherence, we first use \texttt{librosa} \cite{McFee2015librosaAA} to detect beats and employ beat continuity metrics from \cite{Raffel2014MIREVALAT}. Specifically, we use CMLt and AMLt, which measure whether synthesized audio maintains a consistent beat at the correct tempo (CMLt) or integer multiples or divisions of tempo (AMLt).

\vspace{-3mm}
\section{Results}
\label{sec:result}
\vspace{-2mm}
our model's key capabilities are evaluated in this section.
\vspace{-3mm}
\subsection{Temporal Granularity}
\vspace{-1mm}
We train our proposed method with drum MIDI representations of different temporal resolutions.

As expected, the temporal resolution of the input MIDI has a direct impact on synthesis quality. As shown in Table ~\ref{table:granularity}, performance consistently improves when resolution doubles across all metrics. The model conditioned on 64th-note representations achieves the best results for every metric, demonstrating its ability to leverage fine-grained rhythmic detail. 
We therefore use the 64th-note resolution for all subsequent experiments.

\renewcommand{\arraystretch}{1.3}
\begin{table}[t]
\vspace{-5mm}
\caption{Impact of the resolution of MIDI arrangement representations.}
\label{table:granularity}
\centering
\resizebox{\linewidth}{!}{
\begin{tabular}{ccccccc}
\hline
\toprule
\multirow{2}{*}{Timestep Resolution} & \multicolumn{2}{c}{Audio Quality} & \multicolumn{2}{c}{Alignment} & \multicolumn{2}{c}{Beat Continuity} \\ \cline{2-7}
                                      & FAD$_{VGG}$ $\downarrow$& FAD$_{CLAP}$$\downarrow$& F1 $\uparrow$& RMS Err.$\downarrow$& CMLt$\uparrow$& AMLt $\uparrow$\\
\midrule
16th                                  &0.14            &0.071 &58.33      &13.55           &0.34           &0.44    \\
32nd                                  &\underline{0.11}     &\underline{0.065}           &\underline{64.12}&\underline{12.24} &\underline{0.39}            &\underline{0.49}    \\
64th                                  &\textbf{0.09}&\textbf{0.061}  &\textbf{70.08} &\textbf{10.53}   &\textbf{0.42}             &\textbf{0.51}    \\ \hline
\bottomrule
\end{tabular}}
\centering
\end{table}

\vspace{-5mm}
\subsection{Performance on Diverse Input}
\begin{figure}[h]
    \centering
    \vspace{-2mm}
    \includegraphics[width=\columnwidth]{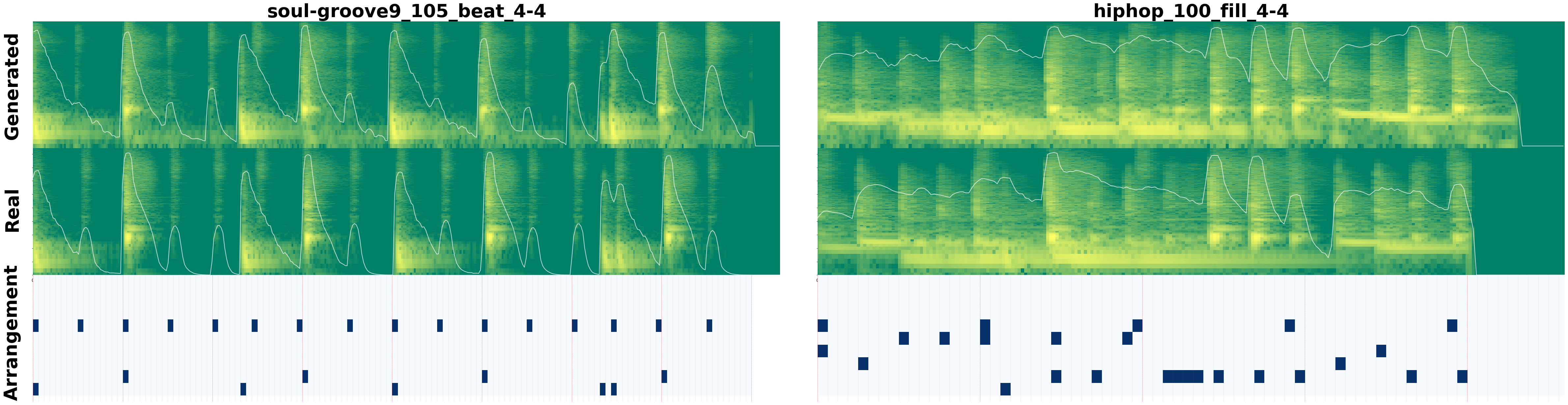}
    \vspace{-5mm}
    \caption{Synthesized drum audio from two types of arrangement MIDI, comparing with ground-truth. Left: 2-bar beat with tempo of 105. Right: 1-bar fill with tempo of 100. Third line visualizes input MIDI representations.}
    \label{fig:beat-vs-fill}
    \vspace{-5mm}
\end{figure}

\vspace{-1mm}
We analyze the model's performance on different subsets of the test data, specifically evaluating its consistency across two arrangement types: (1) beat: repeating rhythmic patterns that form the backbone of a song, and (2) fill: short, often more complex or varied drum phrases used to transition between sections or add excitement. 
As shown in Table ~\ref{table:subsets}, our method works similarly well for both beat and fill patterns. Qualitative results are illustrated in Fig ~\ref{fig:beat-vs-fill}.

The ability to handle both multi-instrument and single-instrument drum patterns is assessed by evaluating our method on the mixture (E-GMD) and stem (StemGMD) respectively.
It is observed that our model tends to have a better performance on StemGMD. This is likely because StemGMD contains single-instrument arrangement, presenting a simpler synthesis task. The difference is obvious in the Alignment metrics, which indicates that the clean hit audio in StemGMD are easier to synthesize than the dense, overlapping hits of E-GMD's mixture drum audio.
Qualitative results are shown in Fig.~\ref{fig:stem}, where our method successfully followed the desired timbre and patterns.
\renewcommand{\arraystretch}{1.3}
\begin{table}[t]
\vspace{-5mm}
\caption{Analysis on diverse rhythm and drum instrument patterns.}
\label{table:subsets}
\centering
\resizebox{\linewidth}{!}{
\begin{tabular}{cccccccc}
\hline
\toprule
\multirow{2}{*}{Set} & \multirow{2}{*}{Input Arrangement Type} & \multicolumn{2}{c}{Audio Quality} & \multicolumn{2}{c}{Alignment} & \multicolumn{2}{c}{Beat Continuity} \\ \cline{3-8}
& & FAD$_{VGG}$ $\downarrow$& FAD$_{CLAP}$$\downarrow$& F1 $\uparrow$& RMS Err.$\downarrow$& CMLt$\uparrow$& AMLt $\uparrow$\\

\midrule
\rowcolor{gray!25}
\multirow{3}{*}{\cellcolor{white}{EGMD}}    &Beat + Fill  &0.18           &\textbf{0.072}           &60.91      &13.32           & \underline{0.43}           & \underline{0.62}  \\   
                         &Beat        &0.28           &0.089           &57.05      &12.89           &\textbf{0.45}            &\textbf{0.69}  \\ &Fill        &0.20           &0.093           &65.31      &13.84           & 0.42            & 0.54   \\ \hline
\rowcolor{gray!25}
\multirow{3}{*}{\cellcolor{white}{StemGMD}} &Beat + Fill  &\underline{0.10}           &\underline{0.073}           &\underline{73.74}      &\underline{9.42}            & 0.41           & 0.47    \\ 
                         &Beat        &0.15           &0.085           &\textbf{74.82}      &9.55            &\underline{0.43}            &0.49     \\
                         &Fill        &\textbf{0.07}  &0.079           &72.08      &\textbf{9.22}            & 0.38            & 0.43   \\ 
\hline
\bottomrule
\end{tabular}}
\centering
\vspace{-2mm}
\end{table}

\renewcommand{\arraystretch}{1.3}
\begin{table}[t]
\vspace{-6mm}
\caption{Impact of conditioning mechanisms and inputs.}
\label{table:ablation}
\centering
\resizebox{0.99\linewidth}{!}{
\begin{tabular}{ccccccccc}
\hline
\toprule
\multirow{2}{*}{Method} & \multirow{2}{*}{Input MIDI Type} & \multirow{2}{*}{Reference MIDI Type} & \multicolumn{2}{c}{Audio Quality} & \multicolumn{2}{c}{Alignment} & \multicolumn{2}{c}{Beat Continuity} \\ \cline{4-9}
                           &    &       & FAD$_{VGG}$ $\downarrow$& FAD$_{CLAP}$$\downarrow$& F1 $\uparrow$& RMS Err.$\downarrow$& CMLt$\uparrow$& AMLt $\uparrow$\\
\midrule
\multirow{3}{*}{Proposed}&Arrangement     & GT Tap   & \textbf{0.09}             &   \textbf{0.061}              &70.08            &\underline{10.53}                  &\underline{0.42}                  &\underline{0.51}                  \\
&Arrangement     & Pseudo Tap   & \underline{0.10}      & \underline{0.063}            &  \underline{70.66}          &   10.61               &  0.41    &   \underline{0.51}               \\
&Tap             & GT Tap   &0.12       &0.070            &68.65            &11.20                  &0.40      &\underline{0.51}                  \\
Proposed (from scratch) & Arrangement   &  GT Tap    & 22.34            & 1.78                & 13.34            & 134.53                 &  0.04                & 0.07                 \\\hline
Cross-attention &\multirow{2}{*}{Arrangement} & GT Tap    &0.12             &0.067                 &45.62            &17.25                  &0.24                  &0.35                  \\
w/o reference context  & & -     &0.13              &0.064                 &\textbf{70.74}            &\textbf{9.63}                  &\textbf{0.43}                  & \textbf{0.52}                 \\ \hline
Proposed &\multirow{2}{*}{Random} &  GT Tap    & 0.83            & 0.256                & 17.73            & 68.44                 &  0.05                & 0.13                 \\
w/o reference context &    &  -    & 1.43            & 0.339                & 19.41            & 66.60                 &  0.06                & 0.13                 \\
\hline
\bottomrule
\end{tabular}}
\centering
\vspace{-4mm}
\end{table}

\subsection{Conditioning Mechanisms}
\vspace{-1mm}
We analyze the influence of different conditioning mechanisms and inputs under a 64th note timestep resolution. The results from our model using random-initialized MIDI input are shown as a lower-bound of each metric.
Since our proposed method utilizes the Ground-Truth (GT) Tap information of reference audio during training, there is the question of whether our model can generalize to the practical use case where GT Tap is not available.
We evaluate our proposed method on both GT Tap and detected Pseudo Tap by \texttt{librosa} to show its generalization.  Table~\ref{table:ablation} shows similar performance when replacing the ground-truth reference tap with a pseudo-tap detected by \texttt{librosa}. Additionally, using Tap instead of Arrangement as input preserves dynamics and rhythm, enabling flexible multi-instrument drum generation without detailed arrangement. 
We also find training a model from scratch results in largely degraded performance, highlighting the critical role of using pre-trained SAO\cite{evans2024stable}.

We validate the effectiveness of our proposed Hybrid Conditioning Mechanism by (1) using cross-attention conditioning; and (2) replacing the reference context $\mathbf{x}_{ref}$ and $\mathbf{c}_{\text{ref}}$ with noisy input and $\mathbf{c}_{\text{tgt}}$. 
First, replacing the input addition with a standard cross-attention mechanism severely harms Alignment and Beat Continuity. This suggests the proposed method is effetive and suitable for preserving temporal structure in this task.
Second, when the reference context is replaced, we observe a slight improvement in Alignment and Beat Continuity, but a degradation in Audio Quality. 
This highlights a trade-off that, without concatenating reference audio latents at the input, the model focuses more on replicating the local rhythmic structure, but less on following the reference audio's timbre even it is provided by cross-attention layers in Content Encoder. 

\vspace{-2mm}
\section{Conclusion}
\vspace{-1mm}
We presented a new method that addresses the task of controllable MIDI-to-drum audio synthesis. By fine-tuning a pre-trained model with proposed content encoder together with hybrid conditioning mechanism, we achieve high-fidelity synthesis that is controllable and robust. Our experiments confirm that a higher input resolution improves quality and validate our conditioning design, which effectively balances rhythmic accuracy with audio quality. This work offers a new tool for creative production, and future work could extend our framework towards automatic composition, by training a model to generate a compatible drum track directly from a given musical piece, such as a melody or a full mix \footnote{Acknowledgment: We thank Marco Martínez and Wei-Hsiang Liao for their valuable feedback.}.

\clearpage
\bibliographystyle{IEEEbib}
{
\footnotesize
\bibliography{strings,refs}
}

\end{document}